\documentclass[conference, 9pt]{IEEEtran}
\IEEEoverridecommandlockouts

\usepackage{cite}
\usepackage{amsmath,amssymb,amsfonts}
\usepackage{algorithmic}
\usepackage{graphicx}
\usepackage{textcomp}
\usepackage{booktabs}
\usepackage{multirow}
\usepackage{xcolor}
\def\BibTeX{{\rm B\kern-.05em{\sc i\kern-.025em b}\kern-.08em
    T\kern-.1667em\lower.7ex\hbox{E}\kern-.125emX}}

\usepackage{fontawesome5}
\usepackage{tikz}
\usetikzlibrary{
    arrows.meta,
    positioning,
    calc,
    fit,
    backgrounds,
    shapes.geometric,
    shadows,
    decorations.pathreplacing
}
\usepackage{fontawesome5}
\usepackage{xcolor}
\usepackage{amsmath}
\usepackage{amsmath,amssymb}
\usepackage{fontawesome5}
\usetikzlibrary{arrows.meta,calc,shapes.geometric,shadows.blur,backgrounds}
\definecolor{lav}{RGB}{237,226,249}
\definecolor{sky}{RGB}{216,235,252}
\definecolor{pale}{RGB}{239,247,255}
\definecolor{pink}{RGB}{255,231,230}
\definecolor{mint}{RGB}{236,248,225}
\definecolor{cream}{RGB}{255,243,216}
\definecolor{purple}{RGB}{244,229,251}
\definecolor{brightblue}{RGB}{13,104,199}
\tikzset{
  panel/.style={draw=black!90,line width=.85pt,rounded corners=5pt,
    top color=white,bottom color=pale,align=center,
    blur shadow={shadow blur steps=4,shadow opacity=.17,shadow xshift=1pt,shadow yshift=-1.5pt}},
  arr/.style={-{Latex[length=2.8mm,width=1.8mm]},line width=1.05pt,draw=black},
  iconblue/.style={text=brightblue},
}

\usepackage{listings}
\usepackage{xcolor}
\usepackage{amsmath}

\lstdefinestyle{rtlExample}{
    basicstyle=\ttfamily\scriptsize,
    columns=fullflexible,
    keepspaces=true,
    breaklines=true,
    breakatwhitespace=false,
    showstringspaces=false,
    frame=single,
    captionpos=b,
    aboveskip=6pt,
    belowskip=6pt
}

\usepackage[most]{tcolorbox}
\usepackage{xcolor}
\usepackage{fontawesome5}
\usepackage{array}

\lstdefinestyle{rtlbase}{
    basicstyle=\ttfamily\scriptsize,
    breaklines=true,
    columns=fullflexible,
    keepspaces=true,
    showstringspaces=false,
    tabsize=2,
    numbers=none,
    frame=none
}

\definecolor{rtlblue}{HTML}{2563A6}
\definecolor{rtlbg}{HTML}{F3F7FC}
\usepackage{cite}
\usepackage{url}
\begin{document}
\bstctlcite{BSTcontrol}
\title{RTLuminate: Illuminating RTL Code Generation with Assertion Feedback\\
}

\author{
    Arash Ardakani$^{\dagger}$,
    Yujing Wang$^{\ddagger}$,
    Ryan Eun$^{\ddagger}$,
    John Wawrzynek$^{\ddagger}$\\[5pt]
    \small
    $^{\dagger}$University of Alberta
    \qquad
    $^{\ddagger}$University of California, Berkeley \vspace{-0.5cm}
}


\maketitle

\begin{abstract}
Open-source small language models (SLMs) enable local deployment, customization, and cost-effective RTL generation while preserving the confidentiality of sensitive hardware designs. Despite these advantages, open-source SLMs generally lag behind frontier proprietary large language models (LLMs) in generating functionally correct RTL. To narrow this performance gap, we introduce RTLuminate, a reference-assisted framework that leverages assertion-based verification feedback to iteratively refine SLM-generated RTL without fine-tuning. RTLuminate employs an open-source assertion mining tool to extract behavioral properties from trusted reference simulation traces. However, the mined assertions and their failure reports are often difficult to interpret and provide limited actionable guidance for RTL refinement. To address this challenge, RTLuminate uses the same open-source SLM employed for RTL generation to interpret assertion failures, summarize the underlying behavioral discrepancies, and then generate actionable feedback for subsequent refinement iterations. We evaluate RTLuminate on two established RTL generation benchmarks, VerilogEval-Human v2 and RTLLM v2.1 using open-source SLMs ranging from $1.3$B to $14$B parameters. With the proposed iterative refinement approach, RTLuminate improves problem pass rates by up $34.6$ percentage points on VerilogEval-Human v2 and by up to $22.0$ percentage points on RTLLM v2.1 across various open-source SLMs. These results demonstrate the potential of assertion-guided feedback to improve the functional correctness of RTL generated by open-source SLMs. 
\end{abstract}

\begin{IEEEkeywords}
RTL Code Generation, Large Language Models (LLMs), Assertion-Guided Refinement, Hardware Verification, Assertion Mining, Simulation-Based Verification, Automated RTL Debugging.
\end{IEEEkeywords}

\section{Introduction}
Artificial intelligence (AI)-driven solutions have demonstrated excellent performance across various tasks, including natural language processing (NLP) \cite{kalyan2021ammus,devlin2019bert,brown2020language} and computer vision (CV) \cite{he2016deep,dosovitskiy2021image}. As such, AI-driven methods have been widely adopted in hardware design automation to address various tasks, including RTL code generation \cite{origen, rtlcoder, gao2024autovcoder, mage}, design verification \cite{zhang2025exploring, sat24, sat25, demotic, recurrent}, hardware accelerator design \cite{fu2023gpt4aigchip, yan2023viability}, and macro placement \cite{mirhoseini,maskplace,chipformer,diffusion}. Among these methods, large language model (LLM)-based approaches for RTL code generation have received significant attention due to their remarkable capabilities in understanding and generating code \cite{verigen,rtlcoder,origen,craftrtl,rtlfixer,
verilogcoder,mage,verireason,correctrtl,revolution,autoverifix, betterv, codev}.

Recent studies have shown that frontier proprietary LLMs, such as GPT and Claude variants \cite{ravindran2026critical, pinckney2025revisiting}, exhibit remarkable proficiency in generating syntactically and functionally correct RTL code. In fact, frontier proprietary LLMs can exceed an $80\%$ pass rate on commonly used benchmarks, such as VerilogEval \cite{verilogeval} and RTLLM \cite{rtllm}, to the extent that developing more challenging benchmarks for evaluating these models has become an emerging research direction \cite{correctrtl}. Despite their superior performance in RTL code generation, frontier proprietary LLMs often raise concerns regarding privacy and cost efficiency \cite{origen}. These models are typically hosted on external cloud services due to their substantial computational and memory requirements, raising concerns regarding the confidentiality of sensitive hardware designs and intellectual property. More precisely, transmitting design specifications, RTL code and simulation test results to third-party services may introduce data privacy and security risks \cite{king2025userprivacy}. Moreover, their usage is associated with inference costs that can accumulate over time, particularly in iterative workflows that require multiple model invocations.

Open-source LLMs offer an attractive alternative to address the aforementioned concerns by enabling local deployment with greater control over sensitive design data and eliminating per-token API or subscription charges. The effectiveness of open-source LLMs largely depends on their model size and the quality of their training data \cite{craftrtl}. For code generation tasks, these models have been further tuned using synthetic instruction-following data through methods such as Self-Instruct \cite{wang-etal-2023-self-instruct}, enhancing their code generation capabilities. For efficient local deployment, open-source small language models (SLMs) are particularly attractive. In this work, we refer to models with fewer than $15$ billion parameters as SLMs since they can be locally deployed with relatively modest computational resources. However, these advantages often come at the cost of significantly lower functional correctness in RTL generation compared with frontier proprietary LLMs. This motivates the development of efficient refinement techniques that improve the RTL generation capabilities of open-source SLMs without relying on externally hosted servers.

There has been several recent studies in literature exploring different techniques to improve the RTL code generation capabilities of open-source LLMs. RTLCoder \cite{rtlcoder} introduced a synthetic RTL instruction-code dataset and a quality-aware training method to fine-tune open-source LLMs. OriGen \cite{origen} used code-to-code data augmentation and a self-reflection mechanism to further improve RTL generation by leveraging compiler feedback to correct syntactic errors. Correct-by-construction synthetic data generation and targeted code repair training were introduced by CraftRTL \cite{craftrtl} to improve the functional correctness of generated RTL. More recently, VeriReason \cite{verireason} enhanced the reasoning and RTL generation capabilities of open-source LLMs by incorporating testbench feedback into reinforcement learning. While these approaches have demonstrated promising results, they require additional training resources and high-quality RTL datasets.

Different from these training-based approaches, we propose RTLuminate to improve the functional correctness of open-source SLMs without additional fine-tuning. RTLuminate was inspired by conventional industrial hardware verification practices where assertions are used to detect functional violations and guide design debugging. RTLuminate used an open-source non-LLM assertion mining tool, called HARM \cite{harm}, to extract behavioral properties from trusted reference RTL simulation traces. The same locally deployed SLM then interprets assertion failures and generates actionable feedback to iteratively refine the generated RTL. This approach incorporates verification-guided debugging into open-source SLM-based RTL generation without relying on proprietary LLMs or additional model training. We evaluate RTLuminate on two well-established benchmarks of VerilogEval-Human v2 \cite{verilogeval} and RTLLM v2.1 \cite{rtllm} using open-source SLMs ranging from $1.3$B to $14$B parameters. Experimental results demonstrate that RTLuminate improves problem pass rates by up to $34.6$ percentage points on VerilogEval-Human v2 and by $22.0$ percentage points on RTLLM v2.1 across various open-source models, highlighting the effectiveness of assertion-guided refinement for improving RTL generation without relying on proprietary LLMs.

\section{Preliminaries}
\label{sec:preliminaries}

\subsection{Feedback-Guided RTL Code Refinement}
\label{sec:feedback_refinement}

Feedback-guided refinement techniques refer to an iterative process in which an SLM first generates an initial implementation and subsequently refines it based on the feedback obtained from compilation, simulation, or other verification tools \cite{rtlfixer,origen,verilogcoder,correctrtl}. Unlike conventional code generation methods that rely on a single SLM inference, these approaches allow the model to first identify errors based on the feedback and then correct them in a subsequent SLM inference. The number of iterations in such refinement frameworks is usually bounded by a fixed number. Moreover, iterative refinement can be applied to multiple independently generated RTL candidates in parallel, allowing each candidate to undergo its own refinement process. In this way the likelihood of obtaining at least one functionally correct implementation among the generated candidates is increased. As a result, this approach leads to improving the pass@$k$ metric, where $k$ denotes the number of independently generated and refined candidates.

Formally, given a design specification $S$ and an SLM $f_{\theta}$ parameterized by $\theta$, the initial RTL code $C^{(0)}$ is generated as
\begin{equation}
    C^{(0)} = f_{\theta}(S).
    \label{eq:initial_generation}
\end{equation}
The generated RTL code is then evaluated using a verification tool $\mathcal{V}$ to obtain feedback $F^{(0)}$, which can be expressed as
\begin{equation}
    F^{(0)} = \mathcal{V}(C^{(0)}, \mathcal{O}),
    \label{eq:verification_feedback}
\end{equation}
where $\mathcal{O}$ represents the verification oracle, such as a testbench or a set of functional/formal properties. At this point, the feedback may contain compilation errors, simulation mismatches, or violated assertions that indicate potential errors in the generated implementation. 

The obtained erroneous feedback is then provided to the SLM to refine the generated RTL. At refinement iteration $t$, this process can be formulated as
\begin{equation}
    C^{(t+1)} =
    f_{\theta}(S, C^{(t)}, F^{(t)}),
    \label{eq:iterative_refinement}
\end{equation}
where $C^{(t)}$ and $F^{(t)}$ represent the generated RTL and its corresponding verification feedback at iteration $t$, respectively. This process continues until the generated RTL satisfies the verification criteria or a predefined maximum number of refinement iterations is reached.

The effectiveness of feedback-guided refinement largely influenced by the quality of the verification feedback and the ability of the SLM to interpret it. While syntactic errors identified by the compiler feedback can be easily interpreted by SLMs, this is not the case for behavioral discrepancies in RTL implementations provided by functional verification feedback. More precisely, raw verification reports may contain limited information about the underlying causes of functional errors, particularly for open-source SLMs with limited reasoning capabilities. This motivates the use of more informative verification feedback to guide the refinement process.

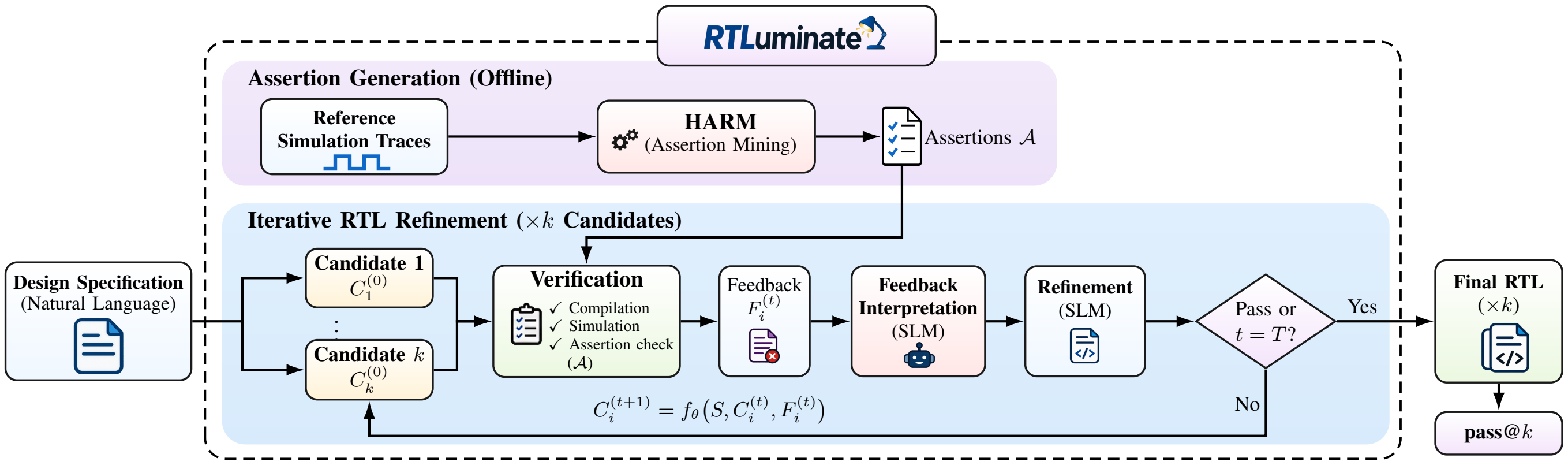
\begin{figure*}[!h]
    \centering
\scalebox{0.75}{
\begin{tikzpicture}[x=1cm,y=1cm,font=\rmfamily]
\draw[dashed,line width=.95pt,rounded corners=9pt,dash pattern=on 5pt off 3pt] (2.78,.20) rectangle (21.05,6.57);
\shade[rounded corners=10pt,top color=lav!70,bottom color=lav] (3.04,4.37) rectangle (15.80,6.28);
\shade[rounded corners=10pt,top color=sky!80,bottom color=sky] (3.04,.42) rectangle (20.88,4.10);
\node[panel,top color=white,bottom color=purple,minimum width=4.25cm,minimum height=.67cm,font=\bfseries\Large] at (11.82,6.66) {\includegraphics[width=3.05cm,trim={4cm 5cm 5cm 5cm},clip]{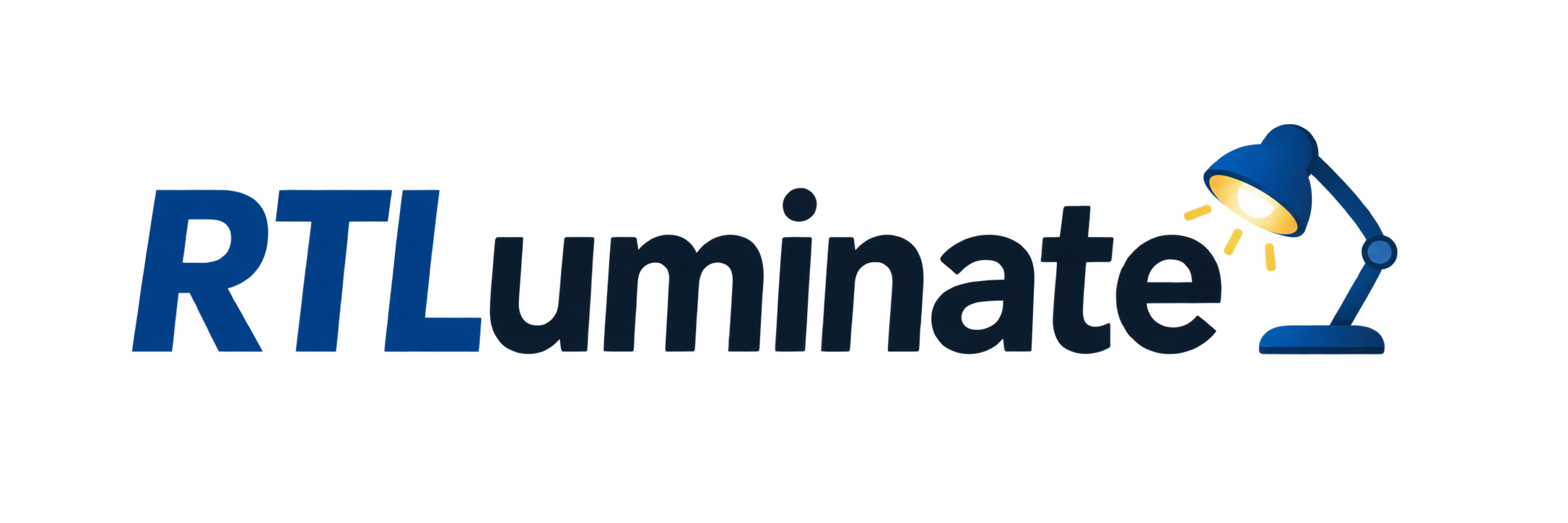}};
\node[anchor=west,font=\bfseries\large] at (3.32,5.99) {Assertion Generation (Offline)};
\node[anchor=west,font=\bfseries\large] at (3.32,3.82) {Iterative RTL Refinement ($\times k$ Candidates)};
\node[panel,minimum width=2.85cm,minimum height=1.15cm] (trace) at (5.06,5.12) {};
\node[align=center,font=\bfseries\small,align=center,text width=2.7cm,anchor=north] at ($(trace.north)+(0,-.08)$) {Reference\\Simulation Traces};
\draw[brightblue,line width=1.6pt] ($(trace.center)+(-.47,-.49)$) -- ++(.18,0) -- ++(0,.19) -- ++(.22,0) -- ++(0,-.19) -- ++(.22,0) -- ++(0,.19) -- ++(.22,0) -- ++(0,-.19) -- ++(.18,0);
\node[panel,top color=white,bottom color=pink,minimum width=3.32cm,minimum height=1.10cm] (harm) at (10.44,5.12) {};
\node[font=\bfseries\large] at ($(harm.center)+(.22,.23)$) {HARM};
\node[font=\small] at ($(harm.center)+(.22,-.15)$) {(Assertion Mining)};
\node[inner sep=0pt]
    at ($(harm.center)+(-1.25,-.04)$)
    {\includegraphics[width=0.65cm]{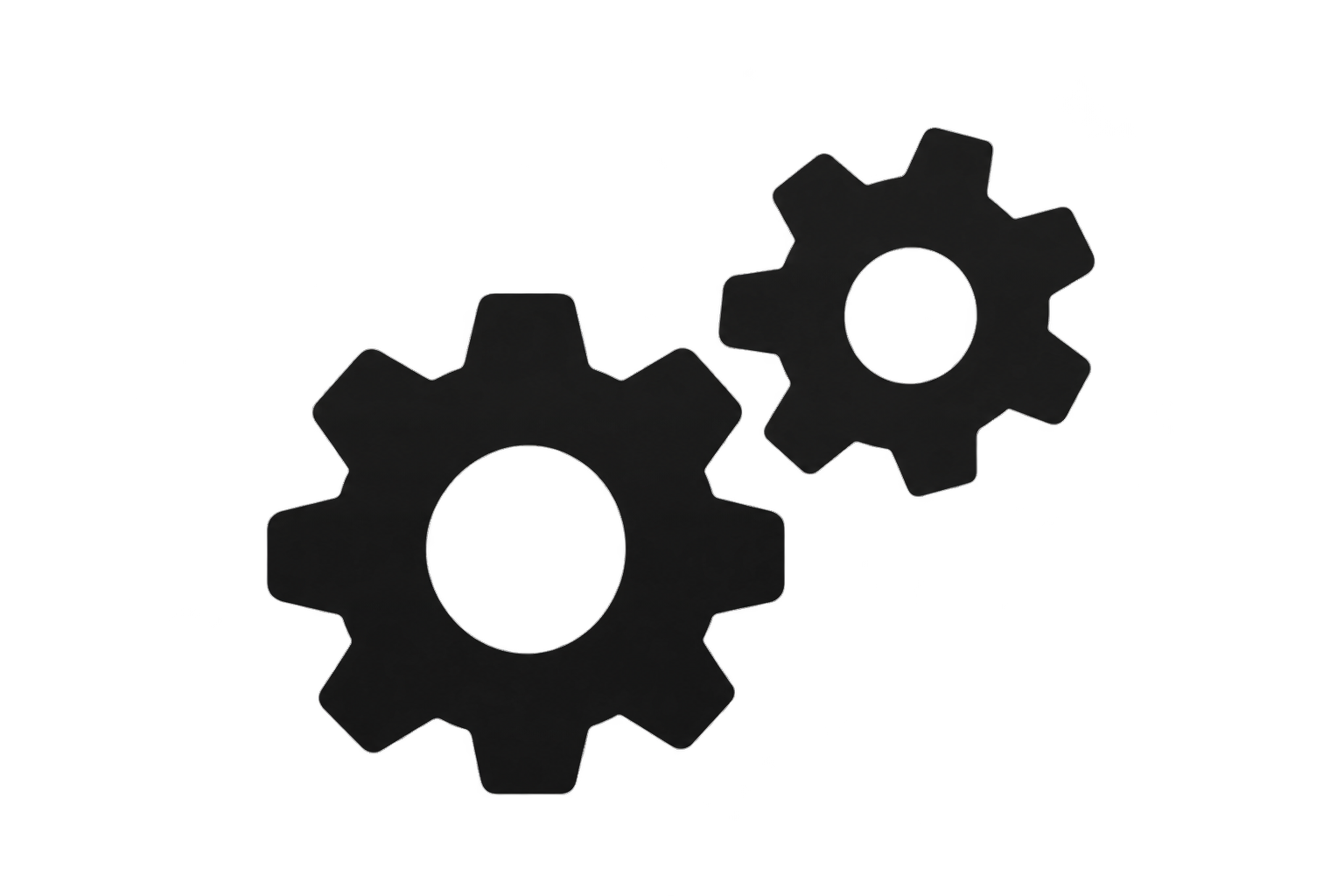}};
\begin{scope}[shift={(13.43,5.12)}]
\draw[fill=white,line width=.95pt,rounded corners=1.1pt] (-.29,-.43) -- (-.29,.44) -- (.09,.44) -- (.29,.24) -- (.29,-.43) -- cycle;
\draw[line width=.75pt] (.09,.44) -- (.09,.24) -- (.29,.24);
\foreach \yy in {.17,-.03,-.23}{\draw[brightblue,line width=1.15pt] (-.20,\yy) -- (-.15,\yy-.05) -- (-.08,\yy+.05);\draw[line width=.75pt] (-.01,\yy) -- (.20,\yy);}
\end{scope}
\node[anchor=west,font=\normalsize] at (13.66,5.12) {Assertions $\mathcal A$};
\draw[arr] (trace.east) -- (harm.west);
\draw[arr] (harm.east) -- (13.11,5.12);
\node[panel,minimum width=2.84cm,minimum height=1.80cm] (spec) at (1.15,2.30) {};
\node[font=\bfseries\small,align=center,anchor=north] at ($(spec.north)+(0,-.10)$) {Design Specification\\[-1pt]\normalfont (Natural Language)};
\node[inner sep=0pt]
    at ($(spec.center)+(0,-0.39)$)
    {\includegraphics[width=1.25cm]{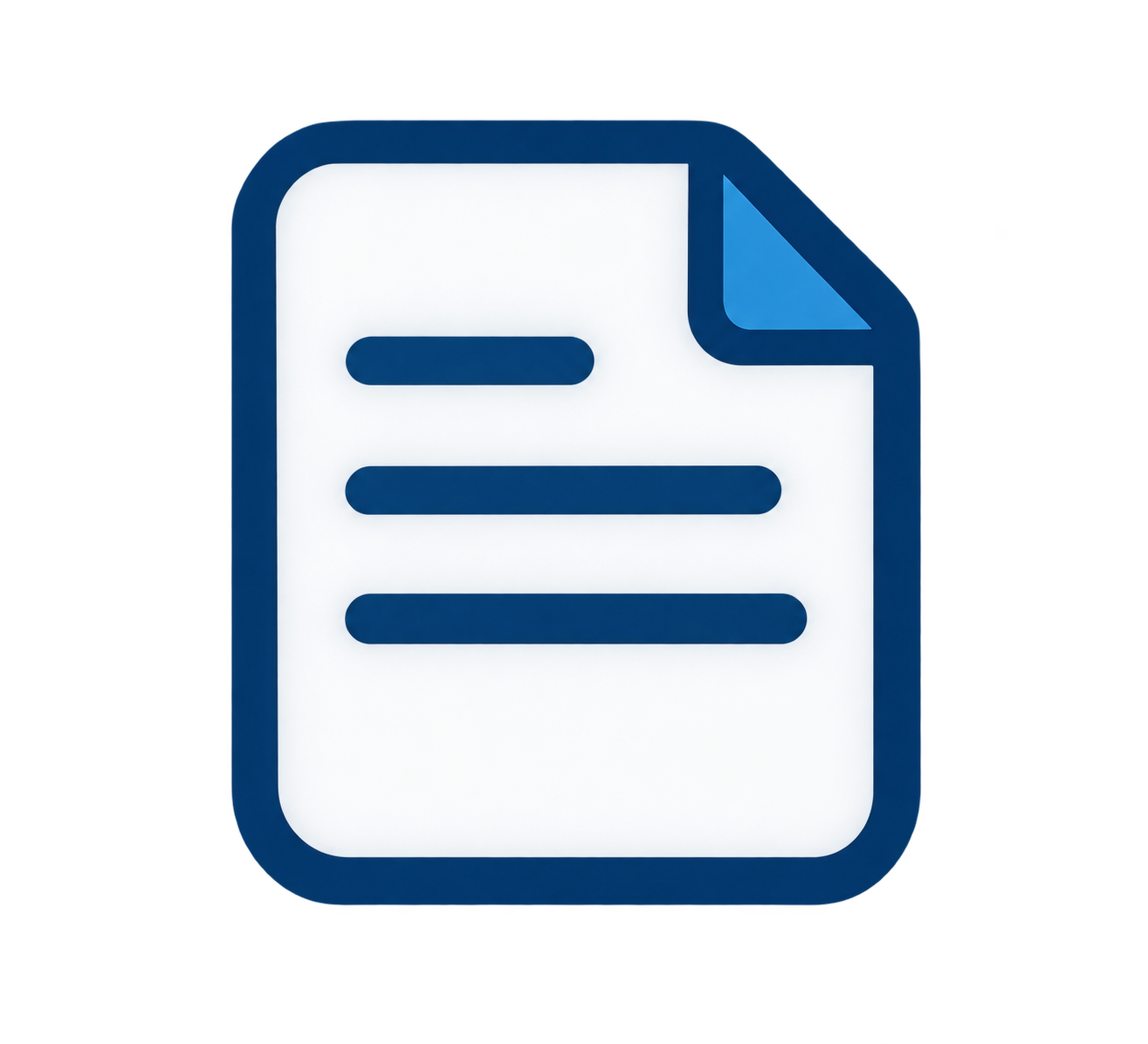}};
\node[panel,top color=white,bottom color=cream,minimum width=1.94cm,minimum height=.85cm] (c1) at (5.29,2.96) {\bfseries Candidate 1\\$C_1^{(0)}$};
\node[font=\large] at (4.79,2.26) {$\vdots$};
\node[panel,top color=white,bottom color=cream,minimum width=1.94cm,minimum height=.85cm] (ck) at (5.29,1.56) {\bfseries Candidate $k$\\$C_k^{(0)}$};
\draw[arr] (spec.east) -- (3.35,2.30) |- (c1.west);
\draw[arr] (3.35,2.30) |- (ck.west);
\node[panel,top color=white,bottom color=mint,minimum width=2.85cm,minimum height=1.70cm] (verify) at (8.61,2.30) {};
\node[font=\bfseries\large] at ($(verify.north)+(0,-.22)$) {Verification};
\node[inner sep=0pt]
    at ($(verify.center)+(-.92,-.04)$)
    {\includegraphics[width=0.65cm]{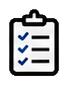}};
\node[anchor=west,align=left,font=\scriptsize] at ($(verify.center)+(-.70,-.23)$) {\checkmark\ Compilation\\\checkmark\ Simulation\\\checkmark\ Assertion check\\\hspace*{1.2em}($\mathcal A$)};
\draw[arr] (c1.east) -- (6.63,2.96) |- (verify.west);
\draw[arr] (ck.east) -- (6.63,1.56) |- (verify.west);
\node[panel,minimum width=1.38cm,minimum height=1.68cm] (feedback) at (11.33,2.30) {};
\node[font=\small,align=center,anchor=north] at ($(feedback.north)+(0,-.08)$) {Feedback\\$F_i^{(t)}$};
\node[font=\large] at ($(feedback.center)+(-.0,-.35)$) {\includegraphics[width=0.65cm]{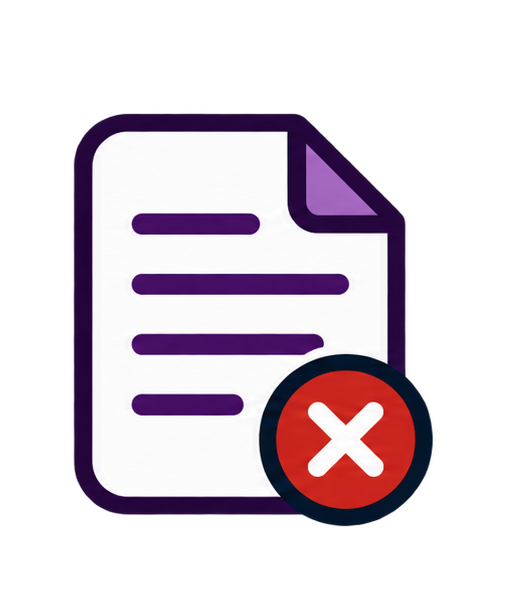}};
\node[panel,top color=white,bottom color=pink,minimum width=2.04cm,minimum height=1.68cm] (interpret) at (13.68,2.30) {};
\node[font=\bfseries\small,align=center,anchor=north] at ($(interpret.north)+(0,-.08)$) {Feedback\\Interpretation\\\normalfont (SLM)};
\node[font=\large,text=brightblue] at ($(interpret.center)+(0,-.49)$) {\includegraphics[width=0.65cm]{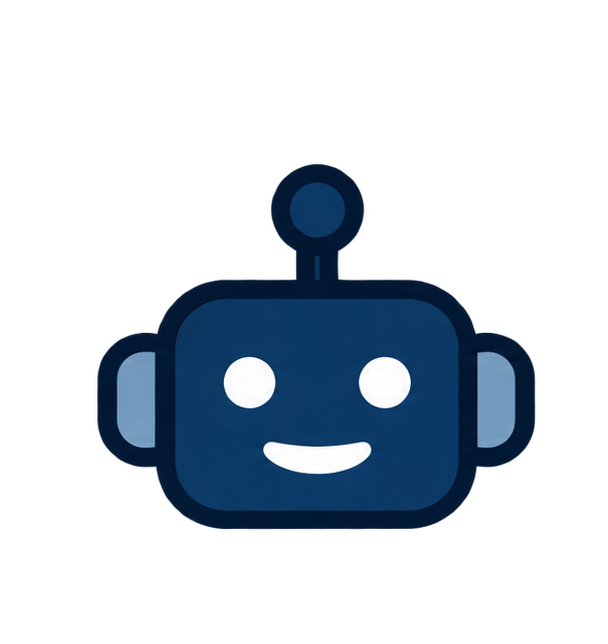}};
\node[panel,minimum width=1.83cm,minimum height=1.68cm] (refine) at (16.23,2.30) {};
\node[font=\bfseries\small,align=center,anchor=north] at ($(refine.north)+(0,-.11)$) {Refinement\\\normalfont (SLM)};
\node[font=\large,text=brightblue] at ($(refine.center)+(0,-.40)$) {\includegraphics[width=0.65cm]{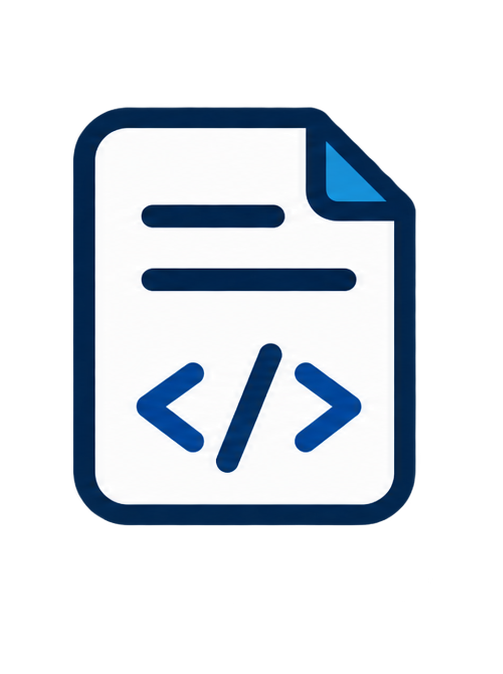}};
\node[diamond,draw=black!90,top color=white,bottom color=purple,aspect=1.48,inner sep=1.8pt,align=center,line width=.85pt] (decision) at (18.99,2.30) {Pass or\\$t=T$?};
\draw[arr] (verify.east) -- (feedback.west);
\draw[arr] (feedback.east) -- (interpret.west);
\draw[arr] (interpret.east) -- (refine.west);
\draw[arr] (refine.east) -- (decision.west);
\draw[arr] (13.43,4.68) -- (13.43,3.58) -- (8.61,3.58) -- (verify.north);
\draw[arr] (decision.south) -- (18.99,.55) -- (5.29,.55) -- (ck.south);
\node[anchor=south,font=\normalsize] at (18.71,.82) {No};
\node[font=\normalsize] at (10.50,0.95) {$C_i^{(t+1)}=f_{\theta}\bigl(S,C_i^{(t)},F_i^{(t)}\bigr)$};
\node[panel,top color=white,bottom color=mint,minimum width=1.94cm,minimum height=1.87cm] (final) at (22.55,2.30) {};
\node[font=\bfseries\small,align=center,anchor=north] at ($(final.north)+(0,-.13)$) {Final RTL\\\normalfont\normalsize ($\times k$)};
\node[font=\fontsize{22}{24}\selectfont] at ($(final.center)+(.13,-.40)$) {\includegraphics[width=1.05cm]{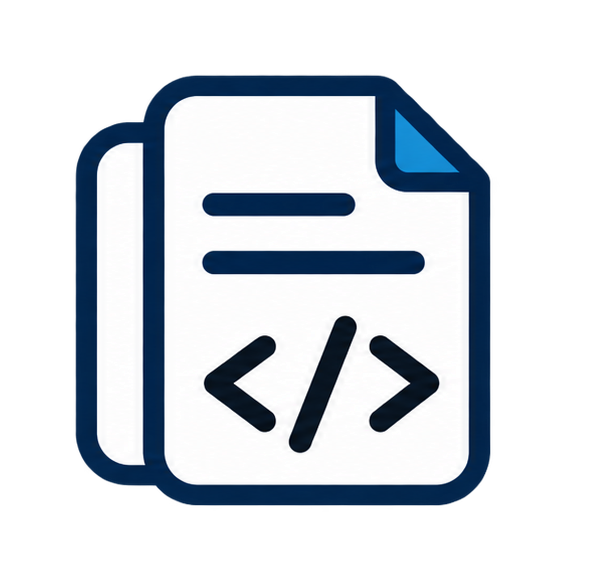}};
\node[panel,top color=white,bottom color=purple,minimum width=1.94cm,minimum height=.65cm,font=\bfseries\normalsize] (pass) at (22.55,.59) {pass@$k$};
\draw[arr] (decision.east) -- node[above left,font=\normalsize]{Yes} (final.west);
\draw[arr] (final.south) -- (pass.north);
\end{tikzpicture}}
    \caption{Overview of the proposed RTLuminate framework. Reference simulation traces are used to automatically extract behavioral assertions using HARM. Given a natural-language specification, an open-source SLM generates multiple RTL candidates, which are evaluated through compilation, simulation, and assertion-based verification. The same SLM interprets verification failures and generates actionable feedback to iteratively refine each candidate until the verification criteria are satisfied or the maximum number of iterations is reached.}
    \label{fig:rtluminate}
    \vspace{-0.15cm}
\end{figure*}

\subsection{Assertion-Based Verification and Mining}
\label{sec:assertion_mining}
Assertion-based verification is a well-known technique in hardware design to verify whether an implementation satisfies a set of predefined functional properties \cite{foster2004assertion}. In particular, assertions describe the expected behavior of a design in the form of logical and temporal relationships among its signals. These properties can be evaluated through simulation or formal verification to identify functional violations. However, manually writing assertions requires in-depth knowledge of the design, making the verification process time-consuming and prone to errors.

To address the aforementioned limitations, several studies have proposed automatic assertion generation techniques to extract functional properties from RTL implementations or their simulation traces \cite{goldmine,aminer,harm}. Assertion mining techniques can be broadly classified into two categories: static and dynamic mining. Static mining approaches analyze the design implementation to extract functional properties, whereas dynamic mining approaches identify behavioral relationships from simulation traces. Among these two categories, dynamic assertion mining is generally more scalable, but is limited to the behaviors observed in the available simulation traces. Nevertheless, dynamic assertion mining is particularly attractive for RTL code generation as the expected design behavior can be obtained directly from reference simulation traces generated by the testbench rather than relying on an LLM to interpret the design specification or RTL code to generate assertions.

Among these approaches, we adopt an open-source non-LLM-based tool, called HARM \cite{harm}, to automatically generate linear temporal logic assertions from simulation traces. Unlike static assertion mining methods, HARM does not require access to the RTL source code and instead relies on observed signal values and a set of user-defined hints to discover temporal relationships.

Let us denote a set of simulation traces as $\mathcal{T} = \{\tau_1,\tau_2,\ldots,\tau_N\}$, where each trace $\tau_i$ consists of a sequence of signal valuations over time. HARM takes the simulation traces and a set of user-defined hints $\mathcal{H}$ as inputs to generate a set of assertions $\mathcal{A}$, expressed as
\begin{equation}
\mathcal{A} = \operatorname{HARM}(\mathcal{T},\mathcal{H}).
\label{eq:harm_mining}
\end{equation}

Although HARM supports user-defined hints as inputs as shown in the above equation, we rely only on simulation traces to generate assertions in this work. HARM performs assertion mining from simulation traces in three main steps. First, HARM instantiates temporal templates using propositions derived from the simulation traces to generate candidate assertions. Second, it evaluates these candidates against the traces and employs an entropy-based decision tree algorithm to identify additional temporal relationships. Finally, the generated assertions are filtered and ranked according to predefined metrics to identify the most relevant behavioral properties. The resulting assertions capture behavioral properties observed in the reference simulation traces. However, due to the finite number of simulation traces, the assertions generated by HARM do not necessarily capture all possible design behaviors. As such, their effectiveness depends on the quality and coverage of the reference testbench.

\section{Methodology}
\label{sec:methodology}

In this section, we introduce RTLuminate, which leverages assertion-based verification feedback to improve the functional correctness of RTL code generated by SLMs. Unlike conventional feedback-guided approaches that directly use raw verification logs for refinement, RTLuminate employs an additional feedback interpretation step to identify functional violations through a set of behavioral assertions and generate actionable guidance for subsequent refinement iterations.

Fig. \ref{fig:rtluminate} illustrates the overall architecture of RTLuminate. The proposed framework consists of two main stages: (1) assertion generation from reference simulation traces and (2) iterative RTL refinement using assertion-based verification feedback. Both stages rely on open-source tools and locally deployed SLMs without requiring additional model training or the use of proprietary LLMs.

\subsection{Assertion Generation}
\label{subsec:offline_assertions}
The first stage of RTLuminate focuses on extracting behavioral properties from reference simulation traces. As discussed in Section \ref{sec:assertion_mining}, manually writing assertions requires in-depth knowledge of the underlying design and verification expertise, making the process time-consuming and less suitable for automated RTL code generation using SLMs. On the other hand, although frontier proprietary LLMs can be employed to generate assertions, their usage is associated with API or subscription costs and may raise security and privacy concerns. Moreover, the assertions generated by these models may not always be reliable due to their probabilistic nature. To eliminate the need for manually written assertions or those generated by frontier proprietary LLMs, we employ HARM, an open-source dynamic assertion mining tool, to automatically generate assertions by extracting temporal properties from reference simulation traces obtained through the testbench.

Let us denote the reference simulation traces as $\mathcal{T}_{\mathrm{ref}}$, where each trace contains the values of the monitored design signals over time. The assertion generation process using HARM can be expressed as
\begin{equation}
    \mathcal{A} =
    \operatorname{HARM}(\mathcal{T}_{\mathrm{ref}}),
    \label{eq:assertion_generation}
\end{equation}
where $\mathcal{A}=\{a_1,a_2,\ldots,a_m\}$ represents the set of mined behavioral assertions. The generated assertions capture temporal relationships among the monitored signals and are used to identify functional violations in the RTL code generated by SLMs. It is worth mentioning that we employ HARM without specifying any design hints or user-defined assertion templates in our implementation. Moreover, since the generated assertions are extracted from the observed signal values in the reference simulation traces, they do not necessarily capture all possible input combinations and design states. As such, their effectiveness largely depends on the coverage of the reference simulation traces.

To better illustrate how RTLuminate improves the functional correctness of SLM-generated RTL code, let us consider an example of a three-input combinational circuit. The circuit consists of three single-bit inputs, $a$, $b$, and $c$, and a single-bit output, $out$. Fig.~\ref{fig:example-spec} presents the original problem specification from VerilogEval.

\begin{figure}[t]
    \centering
    \begin{tcolorbox}[
        enhanced,
        colback=rtlbg,
        colframe=rtlblue,
        colbacktitle=rtlblue,
        coltitle=white,
        title={\faIcon{file-alt}\quad Natural-Language Specification},
        fonttitle=\bfseries\small,
        arc=2mm,
        boxrule=0.6pt,
        left=4pt,
        right=4pt,
        top=4pt,
        bottom=4pt]

    \small
    I would like you to implement a module named
    \texttt{TopModule} with the following interface.
    All input and output ports are one bit unless
    otherwise specified.

    \begin{flushleft}
    \ttfamily
    \hspace{1em}- input  a\\
    \hspace{1em}- input  b\\
    \hspace{1em}- input  c\\
    \hspace{1em}- output out
    \end{flushleft}

    The module should implement the circuit described
    by the Karnaugh map below.

    \begin{center}
        \renewcommand{\arraystretch}{1.2}
        \begin{tabular}{c|cc}
            & \multicolumn{2}{c}{$a$} \\
            $bc$ & $0$ & $1$ \\ \hline
            $00$ & $0$ & $1$ \\
            $01$ & $1$ & $1$ \\
            $11$ & $1$ & $1$ \\
            $10$ & $1$ & $1$
        \end{tabular}
    \end{center}

    \end{tcolorbox}
    \vspace{-0.25cm}
    \caption{Original natural-language specification
    of \texttt{Prob050\_kmap1} from VerilogEval-Human v2.}
    \label{fig:example-spec}
    \vspace{-0.5cm}
\end{figure}

For this problem, HARM generates eight accepted assertions corresponding to four distinct behavioral properties. Fig~\ref{fig:example-harm} presents a compact representation of the generated assertion monitors, where equivalent assertions are grouped together while preserving their original sampling conditions and defined-value guards. It is worth mentioning that the testbench employs a sampling clock for assertion evaluation, which is not part of the combinational circuit.

\begin{figure}[t]
    \centering
    \begin{tcblisting}{
        enhanced,
        listing only,
        listing engine=listings,
        listing options={
            style=rtlbase,
            language=Verilog,
            numbers=none,
            keywordstyle=\color{rtlblue}\bfseries,
            commentstyle=\color{gray},
            basicstyle=\ttfamily\scriptsize
        },
        colback=rtlbg,
        colframe=rtlblue,
        colbacktitle=rtlblue,
        coltitle=white,
        title={\faIcon{shield-alt}\quad HARM-Derived Assertion Monitors},
        fonttitle=\bfseries\small,
        arc=2mm,
        boxrule=0.6pt,
        left=3pt,
        right=3pt,
        top=3pt,
        bottom=3pt
    }
always @(negedge clk) begin
    #1;  // 1 ps settling delay

    if (verification_enabled) begin
        // p_harm_004 and p_harm_005
        if ((a === 1'b1) &&
            !$isunknown(out_dut))
            assert (out_dut === 1'b1)
                else $error("a=1 requires out=1");

        // p_harm_000 and p_harm_001
        if ((b === 1'b1) &&
            !$isunknown(out_dut))
            assert (out_dut === 1'b1)
                else $error("b=1 requires out=1");

        // p_harm_002 and p_harm_003
        if ((c === 1'b1) &&
            !$isunknown(out_dut))
            assert (out_dut === 1'b1)
                else $error("c=1 requires out=1");

        // p_harm_006 and p_harm_007
        if ((a === 1'b0) &&
            (b === 1'b0) &&
            (c === 1'b0) &&
            !$isunknown(out_dut))
            assert (out_dut === 1'b0)
                else $error(
                    "a=b=c=0 requires out=0");
    end
end
    \end{tcblisting}
\vspace{-0.25cm}
    \caption{HARM-derived assertion monitors for
    \texttt{Prob050\_kmap1}. The testbench time unit is 1 ps.}
    \label{fig:example-harm}
\end{figure}

\subsection{Initial RTL Generation}
\label{subsec:initial_generation}
RTLuminate starts the initial RTL code generation by employing an open-source SLM $f_{\theta}$, parameterized by $\theta$, given a natural-language specification $S$, such that
\begin{equation}
    C_i^{(0)} \sim p_{\theta}(C \mid S),
    \qquad i=1,\ldots,k,
    \label{eq:initial_candidates}
\end{equation}
where $C_i^{(0)}$ denotes the $i$-th initial RTL candidate and $p_{\theta}$ represents the conditional probability distribution defined by the SLM. RTLuminate supports generating $k$ independent RTL candidates to increase the probability of generating functionally correct RTL code where the effectiveness of this approach can be evaluated using the pass@$k$ metric. Following our previous example, the SLM generates the initial RTL implementation shown in Fig \ref{fig:example-failed}, which is functionally incorrect.

\begin{figure}[t]
    \centering
    \begin{tcblisting}{
        enhanced,
        listing only,
        listing engine=listings,
        listing options={
            style=rtlbase,
            language=Verilog,
            numbers=none,
            keywordstyle=\color{rtlblue}\bfseries,
    commentstyle=\color{gray},
            basicstyle=\ttfamily\scriptsize,
            literate=*{~}{{\raisebox{-0.75ex}{\textasciitilde}}}1
        },
        colback=rtlbg,
        colframe=rtlblue,
        colbacktitle=rtlblue,
        coltitle=white,
        title={\faFileCode\quad Initial RTL -- Iteration 0},
        fonttitle=\bfseries\small,
        arc=2mm,
        boxrule=0.6pt,
        left=3pt,
        right=3pt,
        top=3pt,
        bottom=3pt
    }
module TopModule (
    input a,
    input b,
    input c,
    output out
);

    // Implement the circuit using the Karnaugh map
    assign out = (a & b) | (~a & ~b & c);

endmodule
    \end{tcblisting}
    \vspace{-0.25cm}
    \caption{Initial RTL implementation generated by
    Qwen2.5-Coder-3B-Instruct (candidate 0, iteration 0).}
    \label{fig:example-failed}
\vspace{-10pt}
\end{figure}

\subsection{Assertion-Based Verification}
\label{subsec:assertion_verification}
Following the initial RTL generation, each candidate is first checked for syntax errors through compilation. Candidates that compile successfully are then passed through simulation and assertion-based verification, where they are evaluated against the expected behavioral properties. Whether the generated RTL code encounters compilation errors, simulation failures, or assertion violations, the resulting verification feedback $F_i^{(t)}$ is obtained for the $i$-th candidate at refinement iteration $t$ as
\begin{equation}
    F_i^{(t)} =
    \mathcal{V}\left(C_i^{(t)},\mathcal{A}\right),
    \label{eq:verification_feedback}
\end{equation}
where $\mathcal{V}$ denotes the verification process. Let us denote the set of assertions violated by the generated RTL code as
\begin{equation}
\begin{aligned}
    \mathcal{A}_{i,\mathrm{fail}}^{(t)}
    = \big\{a_j \in \mathcal{A} \mid
    a_j \text{ is violated by}~C_i^{(t)}\big\}.
\end{aligned}
\label{eq:failed_assertions}
\end{equation}
These violations indicate one or more behavioral properties of the generated RTL code that differ from the expected functional behavior. Fig. \ref{fig:example-raw} presents three representative assertion failure records obtained from the simulation of the initial RTL implementation in our previous example.

\begin{figure}[t]
    \centering
    \begin{tcolorbox}[
        enhanced,
        colback=rtlbg,
        colframe=rtlblue,
        colbacktitle=rtlblue,
        coltitle=white,
        title={\faExclamationTriangle\quad Assertion Verification Failures},
        fonttitle=\bfseries\small,
        arc=2mm,
        boxrule=0.6pt,
        left=3pt,
        right=3pt,
        top=3pt,
        bottom=3pt
    ]
    \ttfamily\tiny
    \raggedright
    \setlength{\parindent}{0pt}
    \sloppy

    \detokenize{ASSERTION_FAILED property=p_harm_000 time=11 requirement=(b equals 1) implies same sample: (out equals 1) observed_consequent=0 trigger_context=a=0 b=1 c=1 out_dut=0 current_context=a=0 b=1 c=1 out_dut=0}

    \medskip

    \detokenize{ASSERTION_FAILED property=p_harm_002 time=11 requirement=(c equals 1) implies same sample: (out equals 1) observed_consequent=0 trigger_context=a=0 b=1 c=1 out_dut=0 current_context=a=0 b=1 c=1 out_dut=0}

    \medskip

    \detokenize{ASSERTION_FAILED property=p_harm_004 time=71 requirement=(a equals 1) implies same sample: (out equals 1) observed_consequent=0 trigger_context=a=1 b=0 c=0 out_dut=0 current_context=a=1 b=0 c=0 out_dut=0}

    \end{tcolorbox}
\vspace{-0.25cm}
    \caption{Selected raw assertion failures from the
    initial RTL implementation.}
    \label{fig:example-raw}
\end{figure}

\subsection{SLM-Based Feedback Interpretation}
\label{subsec:feedback_interpretation}
While compiler errors primarily indicate syntax issues, assertion failures identify functional discrepancies between the expected and observed behaviors. However, raw assertion failure logs may not directly provide explanations regarding the underlying causes of these errors. As such, interpreting simulation-based assertion failure reports can be challenging for SLMs.

To address the aforementioned limitations, we introduce a feedback interpretation step that transforms raw simulation-based assertion failure reports into natural-language summaries of potential root causes for subsequent RTL refinement. More precisely, RTLuminate employs the same open-source SLM to analyze the verification feedback and identify potential root causes of the behavioral discrepancies. The SLM-generated feedback summary $G_i^{(t)}$ can be expressed as
\begin{equation}
    G_i^{(t)} =
    f_{\theta}\left(
    P_{\mathrm{interpret}},
    S,
    C_i^{(t)},
    F_i^{(t)}
    \right),
    \label{eq:feedback_interpretation}
\end{equation}
where $P_{\mathrm{interpret}}$ represents the prompt used for feedback interpretation. This intermediate feedback interpretation step separates error identification from RTL correction, allowing the SLM to focus on understanding the verification failures before generating a revised RTL implementation.

Fig. \ref{fig:example-summary} presents the actual feedback summary generated by the SLM. As shown in Fig. \ref{fig:example-summary}, the SLM successfully interprets the three assertion violations and summarizes them into a set of behavioral requirements. In particular, the generated feedback indicates that the output must be asserted whenever at least one of the three inputs is high.

\begin{figure}[t]
    \centering
    \begin{tcolorbox}[
        enhanced,
        colback=rtlbg,
        colframe=rtlblue,
        colbacktitle=rtlblue,
        coltitle=white,
        title={\faRobot\quad SLM-Generated Feedback},
        fonttitle=\bfseries\small,
        arc=2mm,
        boxrule=0.6pt,
        left=5pt,
        right=5pt,
        top=5pt,
        bottom=5pt
    ]
    \small

    To implement the \texttt{TopModule} according to the
    provided specification and Karnaugh map, follow these rules:

    \begin{itemize}
        \setlength{\itemsep}{2pt}
        \setlength{\topsep}{3pt}
        \setlength{\leftmargin}{1em}
        \item If \texttt{b} is 1, then \texttt{out} must be 1
        at the same sample time.
        \item If \texttt{c} is 1, then \texttt{out} must be 1
        at the same sample time.
        \item If \texttt{a} is 1, then \texttt{out} must be 1
        at the same sample time.
    \end{itemize}

    These rules ensure that the output \texttt{out} is high
    when any of the inputs \texttt{a}, \texttt{b}, or
    \texttt{c} are high, maintaining the logic described
    by the Karnaugh map.

    \end{tcolorbox}
\vspace{-0.25cm}
    \caption{Verbatim feedback summary generated by
    Qwen2.5-Coder-3B-Instruct.}
    \label{fig:example-summary}
    \vspace{-0.5cm}
\end{figure}

\subsection{Iterative RTL Refinement}
\label{subsec:iterative_refinement}
In this step, the SLM is provided with either compilation errors or the SLM-generated assertion feedback summary to refine the RTL code. The refinement process at iteration $t$ can be formulated as
\begin{equation}
    C_i^{(t+1)} =
    f_{\theta}\left(
    P_{\mathrm{refine}},
    S,
    C_i^{(t)},
    G_i^{(t)}
    \right),
    \label{eq:rtl_refinement}
\end{equation}
where $P_{\mathrm{refine}}$ represents the refinement prompt and $C_i^{(t+1)}$ denotes the refined RTL code. The refined code is then compiled, simulated, and evaluated against the generated assertions. If the candidate passes the verification step, the refinement process is terminated and the candidate is considered a valid solution. Otherwise, the verification feedback is analyzed again to generate updated refinement guidance. This process is repeated until either a candidate passes the verification step or the maximum number of refinement iterations $T_{\max}$ is reached. The resulting implementations are evaluated using the pass@$k$ metric, which measures the proportion of problems for which at least one of the $k$ candidates passes the functional evaluation. Following our previous example, after incorporating the summarized assertion feedback, the same candidate generates the refined RTL implementation shown in Fig. \ref{fig:example-correct}, which passes the functional verification.

\begin{figure}[t]
    \centering
    \begin{tcolorbox}[
        enhanced,
        colback=rtlbg,
        colframe=rtlblue,
        colbacktitle=rtlblue,
        coltitle=white,
        title={\faCheckCircle\quad Refined RTL -- Iteration 1},
        fonttitle=\bfseries\small,
        arc=2mm,
        boxrule=0.6pt,
        left=5pt,
        right=5pt,
        top=5pt,
        bottom=5pt
    ]
    \ttfamily\scriptsize
    \noindent
    \textcolor{rtlblue}{\textbf{module}} TopModule (\\
    \hspace*{1em}\textcolor{rtlblue}{\textbf{input}} a,\\
    \hspace*{1em}\textcolor{rtlblue}{\textbf{input}} b,\\
    \hspace*{1em}\textcolor{rtlblue}{\textbf{input}} c,\\
    \hspace*{1em}\textcolor{rtlblue}{\textbf{output}} out\\
    );\\[5pt]
    \hspace*{1em}\textcolor{rtlblue}{\textbf{assign}}
    out = a | b | c;\\[5pt]
    \textcolor{rtlblue}{\textbf{endmodule}}
    \end{tcolorbox}
\vspace{-0.25cm}
    \caption{Correct RTL implementation generated after
    one feedback iteration (candidate 0, iteration 1).}
    \label{fig:example-correct}
    \vspace{-0.5cm}
\end{figure}

\section{Experimental Results}
\subsection{Experimental Setup}
To evaluate the performance of RTLuminate for RTL code generation, we use two established RTL code generation benchmarks, VerilogEval-Human v2 \cite{verilogeval} and RTLLM v2.1 \cite{rtllm}, covering various types of designs, including arithmetic circuits, control logic such as finite-state machines, and data structures. VerilogEval-Human v2 consists of $156$ problems covering a wide range of combinational and sequential designs with different levels of complexity. RTLLM v2.1, on the other hand, comprises $50$ hand-crafted design problems that generally involve more complex hardware specifications and implementation requirements. These two benchmarks allows us to asses the effectiveness of RTLuminate across a wide range of RTL code generation tasks.

Since this paper targets open-source SLMs due to their efficiency and lower privacy risks, we choose different families of SLMs, including Qwen \cite{qwen}, CodeLlama \cite{codellama}, and DeepSeek \cite{deepseek}, with different sizes ranging from $1.5$B to $14$B parameters. We measure the performance of RTLuminate on both benchmarks across different open-source SLMs in terms of the pass@$k$ metric, which evaluates the proportion of problems for which at least one of the $k$ generated RTL candidates is functionally correct. We vary the number of candidates per generation round (i.e., $k$) over $\{1,3,5\}$ and the maximum number of feedback iterations (i.e., $T_{\max}$) over $\{0,5,10\}$. 

In this setup, zero feedback iterations denotes initial generation only, which serves as the baseline. The number of tokens is limited to $2,048$ new tokens per call for RTL code generation, while feedback summarization uses the same model with a limit of $256$ tokens per call. Each problem has a $600$-second wall-clock timeout. We use deterministic decoding for the first candidate and stochastic sampling for additional candidates. The second candidate uses a temperature of $0.55$, while subsequent candidates use $0.80$ with nucleus sampling of $0.95$ in both cases. A problem is counted as passed when a generated candidate passes both the compilation and simulation checks. RTLuminate stops the refinement process once a passing candidate is found or the maximum number of iterations (i.e., $T_{\max}$) is reached. The reported pass rate is the fraction of benchmark problems solved within the configured search budget.

We run all experiments on a single NVIDIA L40S GPU with $48$GB of GPU memory, hosted on a machine with four AMD EPYC 7R13 vCPUs and approximately $31$GiB of system RAM. We use FP16 (i.e., half-precision floating-point format) for inference using PyTorch. The generated RTL code is compiled and simulated using Icarus Verilog. HARM assertions are mined from reference simulation traces and incorporated into the evaluation testbenches.

\begin{table*}[t]
\centering
\caption{Comparison of RTL generation methods on
VerilogEval-Human and RTLLM benchmarks.}
\vspace{-0.1cm}
\label{tab:comparison}
\small
\setlength{\tabcolsep}{4pt}
\renewcommand{\arraystretch}{1.15}
\scalebox{0.79}{
\begin{tabular}{lllc c c c c c}
\toprule
\toprule
\multirow{2}{*}{Benchmark} &
\multirow{2}{*}{Language Model} &
\multirow{2}{*}{Method} &
\multirow{2}{*}{Parameters} &
\multirow{2}{*}{\shortstack{Open-\\Source}} &
\multirow{2}{*}{\shortstack{Fine-\\Tuned}} &
\multirow{2}{*}{Iterations} &
\multicolumn{2}{c}{Pass@$k$ (\%)} \\
\cmidrule(lr){8-9}
& & & & & & &
$k=1$
& $k=5$ \\
\midrule
\midrule
\multirow{20}{*}{\shortstack{VerilogEval-\\Human v2}}
& Qwen2.5-Coder & RTLuminate
& 1.5B & Yes & No & $0^{*}$~/~$5$~/~$10$
& $5.1$~/~$11.5$ ($+6.4\uparrow$)~/~$11.5$ ($+6.4\uparrow$)
& $12.2$~/~$28.2$ ($+16.0\uparrow$)~/~$28.9$ ($+16.7\uparrow$)\\
& Qwen2.5-Coder & RTLuminate
& 3B & Yes & No & $0^{*}$~/~$5$~/~$10$
& $11.5$~/~$28.9$ ($+17.4\uparrow$)~/~$28.9$ ($+17.4\uparrow$)
& $25.6$~/~$51.9$ ($+26.3\uparrow$)~/~$55.8$ ($+30.2\uparrow$) \\
& Qwen2.5-Coder & RTLuminate
& 7B & Yes & No & $0^{*}$~/~$5$~/~$10$
& $34.0$~/~$42.3$ ($+8.3\uparrow$)~/~$42.3$ ($+8.3\uparrow$)
& $46.2$~/~$60.3$ ($+14.1\uparrow$)~/~$61.5$ ($+15.3\uparrow$) \\
& Qwen2.5-Coder & RTLuminate
& 14B & Yes & No & $0^{*}$~/~$5$~/~$10$
& $39.7$~/~$53.2$ ($+13.5\uparrow$)~/~$53.2$ ($+13.5\uparrow$)
& $51.9$~/~$71.2$ ($+19.3\uparrow$)~/~\textbf{$\textbf{71.2}$} ($+19.3\uparrow$)\\
& DeepSeek-coder & RTLuminate
& 1.3B & Yes & No & $0^{*}$~/~$5$~/~$10$
& $1.3$~/~$12.8$ ($+11.5\uparrow$)~/~$12.8$ ($+11.5\uparrow$)
& $9.0$~/~$31.4$ ($+22.4\uparrow$)~/~$32.1$ ($+23.1\uparrow$) \\
& DeepSeek-coder & RTLuminate
& 6.7B & Yes & No & $0^{*}$~/~$5$~/~$10$
& $16.7$~/~$47.4$ ($+30.7\uparrow$)~/~$51.3$ (\textbf{$\textbf{+34.6}\uparrow$})
& $26.9$~/~$53.9$ ($+27\uparrow$)~/~$57.1$ ($+30.2\uparrow$)\\
& CodeLlama & RTLuminate
& 7B & Yes & No & $0^{*}$~/~$5$~/~$10$
& $12.2$~/~$41.7$ ($+29.5\uparrow$)~/~$46.8$ (\textbf{$\textbf{+34.6}\uparrow$})
& $23.7$~/~$47.4$ ($+23.7\uparrow$)~/~$48.7$ ($+25.0\uparrow$)\\
\cmidrule(lr){2-9}
& Qwen2.5-Coder \cite{verireason} & VeriReason
& 1.5B & Yes & Yes & NA & $23.5$ & $26.7$ \\
& Qwen2.5-Coder \cite{verireason} & VeriReason
& 3B & Yes & Yes & NA & $33.2$ & $47.4$ \\
& Qwen2.5-Coder \cite{verireason} & VeriReason
& 7B & Yes & Yes & NA & $47.9$ & $58.4$ \\
& CodeLlama \cite{verireason} & VeriReason
& 7B & Yes & Yes & NA & $27.5$ & $39.9$ \\

& DeepSeek-coder \cite{rtlcoder} & RTLCoder
& 6.7B & Yes & Yes & NA & $41.6$ & $50.1$ \\

& DeepSeek-coder \cite{origen} & OriGen
& 6.7B & Yes & Yes & NA & $54.4$ & $60.1$ \\

& Qwen2.5-Coder \cite{craftrtl} & CraftRTL
& 7B & Yes & Yes & NA & $68.0$ & $72.4$ \\
& DeepSeek-coder \cite{craftrtl} & CraftRTL
& 6.7B & Yes & Yes & NA & $65.4$ & $70.0$ \\
& CodeLlama \cite{craftrtl} & CraftRTL
& 7B & Yes & Yes & NA & $63.1$ & $67.8$ \\
\cmidrule(lr){2-9}
& GPT-4o \cite{craftrtl} & Proprietary LLM
& NA & No & NA & NA & $57.1$ & $63.9$ \\
& Claude Sonnet 4.5 \cite{craftrtl} & Proprietary LLM
& NA & No & NA & NA & $79.5$ & -- \\
\midrule
\midrule
\multirow{8}{*}{RTLLM v2.1}
& Qwen2.5-Coder & RTLuminate
& 7B & Yes & No & $0^{*}$~/~$5$
& $18.0$~/~$32.0$ ($+14.0\uparrow$)
& $34.0$~/~$56.0$ (\textbf{$\textbf{+22.0}\uparrow$}) \\

& Qwen2.5-Coder & RTLuminate
& 14B & Yes & No & $0^{*}$~/~$5$
& $44.0$~/~$54.0$ ($+10.0\uparrow$)
& $54.0$~/~\textbf{$\textbf{68.0}$} ($+14.0\uparrow$) \\

& DeepSeek-coder & RTLuminate
& 1.3B & Yes & No & $0^{*}$~/~$5$
& 2.0~/~18.0 ($+16.0\uparrow$) & 18.0~/~32.0  ($+14.0\uparrow$)\\

& DeepSeek-coder & RTLuminate
& 6.7B & Yes & No & $0^{*}$~/~$5$
& $28.0$~/~$40.0$ ($+12.0\uparrow$)
& $42.0$~/~$54.0$ ($+12.0\uparrow$) \\

& CodeLlama & RTLuminate
& 7B & Yes & No & $0^{*}$~/~$5$
& $10.0$~/~$20.0$ ($+10.0\uparrow$)
& $20.0$~/~$36.0$ ($+16.0\uparrow$) \\

\cmidrule(lr){2-9}
& GPT-4o \cite{craftrtl} & Proprietary LLM
& NA & No & NA & NA & $56.5$ & $70.3$ \\
& Claude Sonnet 4.5 \cite{craftrtl} & Proprietary LLM
& NA & No & NA & NA & $74.0$ & -- \\
\bottomrule
\bottomrule
\end{tabular}}
\vspace{20pt}
\begin{minipage}{\textwidth}
\footnotesize
\vspace{2pt}
$^{*}$ Iteration 0 denotes the baseline without refinement. Slash-separated results correspond to the refinement iterations listed in the Iterations column in the same order. Values in parentheses indicate improvements over the baseline
in percentage points.
\end{minipage}
\vspace{-1cm}
\end{table*}

\subsection{Key Results}
Table~\ref{tab:comparison} presents the pass rate performance of RTLuminate on both VerilogEval-Human v2 and RTLLM v2.1 benchmarks. The results are reported in the order of refinement iterations with improvements over the baseline shown in parentheses in percentage points. On both benchmarks, RTLuminate consistently improves the baseline pass rate across various open-source SLMs with different architectures and model sizes through its assertion-based feedback refinement process. More specifically, RTLuminate improves the baseline pass rate by up to $34.6$ percentage points on VerilogEval-Human v2 and by up to $22.0$ percentage points on RTLLM v2.1. These improvements demonstrate the effectiveness of assertion-based feedback in improving the functional correctness of SLM-generated RTL code.

In addition to improving the baseline performance, RTLuminate achieves competitive results compared with several Verilog-specific models that have been fine-tuned for RTL code generation. In particular, RTLuminate outperforms VeriReason and RTLCoder using the same underlying SLMs, while achieving competitive performance with OriGen and falling short of CraftRTL. It is worth mentioning that, unlike these approaches, RTLuminate does not require any additional fine-tuning and instead improves RTL generation through an iterative assertion-based feedback refinement process. These results demonstrate that incorporating assertion-based feedback into the RTL generation process can significantly improve the performance of open-source SLMs without modifying their model parameters.

Compared with frontier proprietary LLMs, RTLuminate also narrows the performance gap, although the proprietary models generally achieve higher pass rates. For instance, RTLuminate achieves a pass@$5$ of $71.2\%$ using Qwen2.5-Coder-14B on VerilogEval-Human v2, compared with $63.9\%$ for GPT-4o. On RTLLM v2.1, RTLuminate improves the pass@$5$ of Qwen2.5-Coder-14B from $54.0\%$ to $68.0\%$, approaching the performance of GPT-4o. These results highlight the potential of assertion-based feedback refinement in improving the functional correctness of locally deployed open-source SLMs.

It is worth mentioning that the results of prior fine-tuned RTL generation methods are reported only for VerilogEval-Human v2 in Table~\ref{tab:comparison} as their reported evaluations do not cover RTLLM v2.1. Since earlier versions of RTLLM contain different numbers of problems, directly comparing results across different benchmark versions may not provide a fair evaluation.

\subsection{Ablation Study}
To illustrate the effectiveness of feedback summarization in our framework, we disable some features of RTLuminate and report the results in this section. We compare four configurations using Qwen2.5-Coder-1.5B and Qwen2.5-Coder-3B as shown in Table~\ref{tab:ablation}. In the no-feedback configuration, the model undergoes the refinement process using only the original design specification without any verification feedback. At each refinement iteration, the model is informed that the previous implementation was incorrect and is asked to regenerate a new design based on the original specification. Of course iteratively generating new RTL implementations increases the probability of obtaining functionally correct RTL code in this configuration. In the simulation feedback configuration, we provide all reported non-HARM simulation failures and compiler diagnostics to guide the refinement process. In the HARM-only configuration, we provide unsummarized rules extracted from all HARM assertion failures. Finally, the fourth configuration represents the full RTLuminate framework which incorporates SLM-generated summaries of assertion failures into the refinement process. All configurations are evaluated on the VerilogEval-Human v2 benchmark with a maximum of five refinement iterations. We evaluate the performance of these configurations using five candidates and report the results in terms of the pass rate metric as shown in Table~\ref{tab:ablation}.

\begin{table}[t]
    \centering
    \caption{Ablation results on VerilogEval-Human v2 benchmark.}
    \label{tab:ablation}
    \vspace{-0.1cm}
    \begin{tabular}{lcc}
        \toprule
        \toprule
        Configuration & Qwen2.5-Coder-1.5B & Qwen2.5-Coder-3B \\
        \midrule
        \midrule
        No-feedback
            & $16.7$\% & $39.1$\% \\
        Simulation feedback
            & $21.8$\% & $44.9$\% \\
        HARM-only feedback
            & $20.5$\% & $43.6$\% \\
        RTLuminate
            & $28.2$\% & $51.9$\% \\
        \bottomrule
        \bottomrule
    \end{tabular}
    \vspace{-0.6cm}
\end{table}

As shown in Table~\ref{tab:ablation}, while the simulation feedback and HARM-only configurations improve the pass rate compared with the no-feedback configuration, their improvements are relatively limited. This may be attributed to the difficulty of interpreting raw feedback from the simulator or HARM. Interestingly, the HARM-only configuration achieves a lower pass rate than the simulation feedback configuration, indicating that providing unsummarized assertion rules does not necessarily lead to more effective RTL refinement. In contrast, RTLuminate employs the SLM to interpret assertion failures and generate actionable feedback, resulting in a higher pass rate. These results demonstrate the effectiveness of feedback summarization in improving the functional correctness of generated RTL implementations.

\section{Conclusion}
In this paper, we proposed RTLuminate, an assertion-based feedback refinement framework to improve the functional correctness of RTL code generated by open-source SLMs. RTLuminate employs an open-source assertion mining tool to automatically extract behavioral properties from reference simulation traces without requiring manually written assertions or proprietary LLMs. The generated assertions are then used to identify functional violations in SLM-generated RTL code while the same SLM interprets assertion failures and generates actionable feedback for subsequent refinement iterations. Unlike existing fine-tuning-based approaches, RTLuminate improves RTL generation without requiring additional model training. We evaluated RTLuminate on two established RTL code generation benchmarks, of VerilogEval-Human v2 and RTLLM v2.1, using open-source SLMs ranging from $1.3$B to $14$B parameters. Experimental results demonstrated that RTLuminate improves the baseline pass rate by up to $34.6$ percentage points on VerilogEval-Human v2 and by up to $22.0$ percentage points on RTLLM v2.1. Moreover, RTLuminate achieves competitive performance compared with several fine-tuned RTL generation models and narrows the performance gap with frontier proprietary LLMs. These results demonstrate the effectiveness of assertion-based feedback refinement in improving RTL generation using locally deployed open-source SLMs without additional fine-tuning.

\clearpage
\bibliographystyle{IEEEtran}
\bibliography{DAC}

\begin{thebibliography}{10}
\providecommand{\url}[1]{#1}
\csname url@samestyle\endcsname
\providecommand{\newblock}{\relax}
\providecommand{\bibinfo}[2]{#2}
\providecommand{\BIBentrySTDinterwordspacing}{\spaceskip=0pt\relax}
\providecommand{\BIBentryALTinterwordstretchfactor}{4}
\providecommand{\BIBentryALTinterwordspacing}{\spaceskip=\fontdimen2\font plus
\BIBentryALTinterwordstretchfactor\fontdimen3\font minus \fontdimen4\font\relax}
\providecommand{\BIBforeignlanguage}[2]{{%
\expandafter\ifx\csname l@#1\endcsname\relax
\typeout{** WARNING: IEEEtran.bst: No hyphenation pattern has been}%
\typeout{** loaded for the language `#1'. Using the pattern for}%
\typeout{** the default language instead.}%
\else
\language=\csname l@#1\endcsname
\fi
#2}}
\providecommand{\BIBdecl}{\relax}
\BIBdecl

\bibitem{kalyan2021ammus}
K.~S. Kalyan, A.~Rajasekharan, and S.~Sangeetha, ``Ammus: A survey of transformer-based pretrained models in natural language processing,'' \emph{arXiv preprint arXiv:2108.05542}, 2021.

\bibitem{devlin2019bert}
J.~Devlin, M.-W. Chang, K.~Lee, and K.~Toutanova, ``{BERT}: Pre-training of deep bidirectional transformers for language understanding,'' in \emph{Proceedings of NAACL-HLT}, 2019, pp. 4171--4186.

\bibitem{brown2020language}
T.~B. Brown \emph{et~al.}, ``Language models are few-shot learners,'' in \emph{Advances in Neural Information Processing Systems}, vol.~33, 2020, pp. 1877--1901.

\bibitem{he2016deep}
K.~He, X.~Zhang, S.~Ren, and J.~Sun, ``Deep residual learning for image recognition,'' in \emph{Proceedings of the IEEE Conference on Computer Vision and Pattern Recognition}, 2016, pp. 770--778.

\bibitem{dosovitskiy2021image}
A.~Dosovitskiy \emph{et~al.}, ``An image is worth 16x16 words: Transformers for image recognition at scale,'' in \emph{International Conference on Learning Representations}, 2021.

\bibitem{origen}
F.~Cui \emph{et~al.}, ``{OriGen}: Enhancing rtl code generation with code-to-code augmentation and self-reflection,'' in \emph{Proceedings of the IEEE/ACM International Conference on Computer-Aided Design (ICCAD)}, 2024.

\bibitem{rtlcoder}
S.~Liu, W.~Fang, Y.~Lu, J.~Wang, Q.~Zhang, H.~Zhang, and Z.~Xie, ``{RTLCoder}: Fully open-source and efficient llm-assisted rtl code generation technique,'' \emph{IEEE Transactions on Computer-Aided Design of Integrated Circuits and Systems}, vol.~44, no.~4, pp. 1448--1461, 2025.

\bibitem{gao2024autovcoder}
M.~Gao, J.~Zhao, Z.~Lin, W.~Ding, X.~Hou, Y.~Feng, C.~Li, and M.~Guo, ``Autovcoder: A systematic framework for automated verilog code generation using llms,'' in \emph{2024 IEEE 42nd International Conference on Computer Design (ICCD)}.\hskip 1em plus 0.5em minus 0.4em\relax IEEE, 2024, pp. 162--169.

\bibitem{mage}
Y.~Zhao, H.~Zhang, H.~Huang, Z.~Yu, and J.~Zhao, ``{MAGE}: A multi-agent engine for automated rtl code generation,'' in \emph{Proceedings of the 62nd ACM/IEEE Design Automation Conference (DAC)}, 2025, pp. 1--7.

\bibitem{zhang2025exploring}
Q.~Zhang, W.~Sun, C.~Fang, B.~Yu, H.~Li, M.~Yan, J.~Zhou, and Z.~Chen, ``Exploring automated assertion generation via large language models,'' \emph{ACM Transactions on Software Engineering and Methodology}, vol.~34, no.~3, pp. 1--25, 2025.

\bibitem{sat24}
A.~Ardakani, M.~Kang, K.~He, V.~M. Iyer, S.~Moon, and J.~Wawrzynek, ``Late breaking results: Differential and massively parallel sampling of {SAT} formulas,'' in \emph{Proceedings of the 61st ACM/IEEE Design Automation Conference (DAC)}, 2024, pp. 356:1--356:2.

\bibitem{sat25}
A.~Ardakani, M.~Kang, K.~He, Q.~Huang, and J.~Wawrzynek, ``High-throughput {SAT} sampling,'' in \emph{Proceedings of the Design, Automation \& Test in Europe Conference (DATE)}, 2025.

\bibitem{demotic}
A.~Ardakani, M.~Kang, K.~He, Q.~Huang, V.~M. Iyer, S.~Moon, and J.~Wawrzynek, ``{DEMOTIC}: A differentiable sampler for multi-level digital circuits,'' in \emph{Proceedings of the 30th Asia and South Pacific Design Automation Conference (ASP-DAC)}, 2025, pp. 1328--1335.

\bibitem{recurrent}
A.~Ardakani, K.~He, and J.~Wawrzynek, ``Recurrent {CircuitSAT} sampling for sequential circuits,'' in \emph{Proceedings of the 7th ACM/IEEE Symposium on Machine Learning for CAD (MLCAD)}, 2025.

\bibitem{fu2023gpt4aigchip}
Y.~Fu, Y.~Zhang, Z.~Yu, S.~Li, Z.~Ye, C.~Li, C.~Wan, and Y.~Lin, ``{GPT4AIGChip}: Towards next-generation {AI} accelerator design automation via large language models,'' in \emph{2023 IEEE/ACM International Conference on Computer Aided Design (ICCAD)}, 2023.

\bibitem{yan2023viability}
Z.~Yan, Y.~Qin, X.~S. Hu, and Y.~Shi, ``On the viability of using {LLMs} for {SW/HW} co-design: An example in designing {CiM} {DNN} accelerators,'' in \emph{2023 IEEE 36th International System-on-Chip Conference}, 2023.

\bibitem{mirhoseini}
A.~Mirhoseini \emph{et~al.}, ``{A Graph Placement Methodology for Fast Chip Design},'' \emph{Nature}, vol. 594, no. 7862, pp. 207--212, 2021.

\bibitem{maskplace}
Y.~Lai, Y.~Mu, and P.~Luo, ``{MaskPlace: Fast Chip Placement via Reinforced Visual Representation Learning},'' in \emph{Advances in Neural Information Processing Systems}, 2022.

\bibitem{chipformer}
Y.~Lai, J.~Liu, Z.~Tang, B.~Wang, J.~Hao, and P.~Luo, ``{ChiPFormer: Transferable Chip Placement via Offline Decision Transformer},'' in \emph{Proceedings of the International Conf. on Machine Learning}, 2023.

\bibitem{diffusion}
V.~Lee, M.~Nguyen, L.~Elzeiny, C.~Deng, P.~Abbeel, and J.~Wawrzynek, ``{Chip Placement with Diffusion Models},'' in \emph{Proceedings of the International Conference on Machine Learning}, ser. Proceedings of Machine Learning Research, vol. 267, 2025, pp. 33\,499--33\,514.

\bibitem{verigen}
S.~Thakur \emph{et~al.}, ``{VeriGen}: A large language model for verilog code generation,'' \emph{ACM Transactions on Design Automation of Electronic Systems}, vol.~29, no.~3, 2024, article 46.

\bibitem{craftrtl}
M.~Liu, Y.-D. Tsai, W.~Zhou, and H.~Ren, ``{CraftRTL}: High-quality synthetic data generation for verilog code models with correct-by-construction non-textual representations and targeted code repair,'' in \emph{International Conference on Learning Representations (ICLR)}, 2025.

\bibitem{rtlfixer}
Y.-D. Tsai, M.~Liu, and H.~Ren, ``{RTLFixer}: Automatically fixing rtl syntax errors with large language models,'' in \emph{Proceedings of the 61st ACM/IEEE Design Automation Conference (DAC)}, 2024.

\bibitem{verilogcoder}
C.-T. Ho, H.~Ren, and B.~Khailany, ``{VerilogCoder}: Autonomous verilog coding agents with graph-based planning and abstract syntax tree (ast)-based waveform tracing tool,'' in \emph{Proceedings of the AAAI Conference on Artificial Intelligence}, vol.~39, 2025, pp. 300--307.

\bibitem{verireason}
Y.~Wang, G.~Sun, W.~Ye, G.~Qu, and A.~Li, ``{VeriReason}: Reinforcement learning with testbench feedback for reasoning-enhanced verilog generation,'' in \emph{Proceedings of the Great Lakes Symposium on VLSI (GLSVLSI)}, 2026, pp. 948--953.

\bibitem{correctrtl}
J.~Wang, S.~Liu, H.~Zhou, W.~Fang, Y.~Wu, Y.~Zhu, and Z.~Xie, ``{CorrectRTL}: Boost rtl generation quality through automated self-correction with instruction decomposition,'' in \emph{Proceedings of the IEEE/ACM International Conference on Computer-Aided Design (ICCAD)}, 2026.

\bibitem{revolution}
K.~Min, K.~Cho, J.~Jang, and S.~Kang, ``{REvolution}: An evolutionary framework for rtl generation driven by large language models,'' in \emph{Proceedings of the 31st Asia and South Pacific Design Automation Conference (ASP-DAC)}, 2026.

\bibitem{autoverifix}
Y.~Tan, X.~Meng, Z.~Jiang, and Y.~Lyu, ``{AutoVeriFix}: Automatically correcting errors and enhancing functional correctness in llm-generated verilog code,'' in \emph{Proceedings of the 31st Asia and South Pacific Design Automation Conference (ASP-DAC)}, 2026.

\bibitem{betterv}
Z.~Pei, H.-L. Zhen, M.~Yuan, Y.~Huang, and B.~Yu, ``{BetterV}: Controlled verilog generation with discriminative guidance,'' in \emph{Proceedings of the 41st International Conference on Machine Learning (ICML)}, ser. Proceedings of Machine Learning Research, vol. 235.\hskip 1em plus 0.5em minus 0.4em\relax PMLR, 2024, pp. 40\,145--40\,153.

\bibitem{codev}
Y.~Zhao \emph{et~al.}, ``{CodeV}: Empowering {LLMs} with {HDL} generation through multi-level summarization,'' \emph{arXiv preprint arXiv:2407.10424}, 2024.

\bibitem{ravindran2026critical}
A.~Ravindran, A.~Patra, V.~Babaey, and S.~Purini, ``A critical review and evaluation of {LLMs} for {RTL} generation,'' \emph{IEEE Access}, vol.~14, pp. 28\,522--28\,539, 2026.

\bibitem{pinckney2025revisiting}
N.~Pinckney, C.~Batten, M.~Liu, H.~Ren, and B.~Khailany, ``Revisiting {VerilogEval}: A year of improvements in large-language models for hardware code generation,'' \emph{ACM Transactions on Design Automation of Electronic Systems}, 2025.

\bibitem{verilogeval}
M.~Liu, N.~R. Pinckney, B.~Khailany, and H.~Ren, ``{VerilogEval}: Evaluating large language models for verilog code generation,'' in \emph{Proceedings of the IEEE/ACM International Conference on Computer-Aided Design (ICCAD)}, 2023.

\bibitem{rtllm}
Y.~Lu, S.~Liu, Q.~Zhang, and Z.~Xie, ``{RTLLM}: An open-source benchmark for design rtl generation with large language model,'' in \emph{Proceedings of the Asia and South Pacific Design Automation Conference (ASP-DAC)}, 2024, pp. 722--727.

\bibitem{king2025userprivacy}
J.~King, K.~Klyman, E.~Capstick, T.~Saade, and V.~Hsieh, ``User privacy and large language models: An analysis of frontier developers' privacy policies,'' in \emph{Proceedings of the Eighth AAAI/ACM Conference on AI, Ethics, and Society (AIES)}, vol.~8, no.~2, 2025, pp. 1465--1477.

\bibitem{wang-etal-2023-self-instruct}
Y.~Wang, Y.~Kordi, S.~Mishra, A.~Liu, N.~A. Smith, D.~Khashabi, and H.~Hajishirzi, ``Self-instruct: Aligning language models with self-generated instructions,'' in \emph{Proceedings of the 61st Annual Meeting of the {A}ssociation for {C}omputational {L}inguistics (Volume 1: Long Papers)}, Toronto, Canada, Jul. 2023, pp. 13\,484--13\,508.

\bibitem{harm}
S.~Germiniani and G.~Pravadelli, ``Harm: A hint-based assertion miner,'' \emph{Trans. Comp.-Aided Des. Integ. Cir. Sys.}, vol.~41, no.~11, p. 4277–4288, Nov. 2022.

\bibitem{foster2004assertion}
H.~D. Foster, A.~C. Krolnik, and D.~J. Lacey, \emph{Assertion-Based Design}, 2nd~ed.\hskip 1em plus 0.5em minus 0.4em\relax Springer, 2004.

\bibitem{goldmine}
S.~Vasudevan \emph{et~al.}, ``{GoldMine}: Automatic assertion generation using data mining and static analysis,'' in \emph{Proceedings of the Design, Automation and Test in Europe Conference (DATE)}, 2010.

\bibitem{aminer}
W.~Li \emph{et~al.}, ``{A-Miner}: Automatic mining of assertions from simulation traces,'' in \emph{Proceedings of the International Conference on Computer-Aided Design (ICCAD)}, 2010.

\bibitem{qwen}
J.~Bai \emph{et~al.}, ``Qwen technical report,'' \emph{arXiv preprint arXiv:2309.16609}, 2023.

\bibitem{codellama}
B.~Rozi{\`e}re, J.~Gehring \emph{et~al.}, ``Code llama: Open foundation models for code,'' \emph{arXiv preprint arXiv:2308.12950}, 2023.

\bibitem{deepseek}
D.~Guo \emph{et~al.}, ``Deepseek-coder: When the large language model meets programming -- the rise of code intelligence,'' \emph{arXiv preprint arXiv:2401.14196}, 2024.

\end{thebibliography}

\end{document}